\documentclass[conference]{IEEEtran}

\ifCLASSINFOpdf
\else
\fi
\usepackage{amssymb}
\usepackage{amsmath,amsfonts}
\usepackage{caption2}
\usepackage{color,graphicx}
\usepackage{epsfig}
\usepackage{subfigure}
\usepackage{setspace}
\usepackage{cite}
\usepackage{enumerate}
\usepackage{algorithmic}
\usepackage{algorithm}
\usepackage{array}
\usepackage{multirow}
\usepackage{epstopdf}
\usepackage{dblfloatfix}
\usepackage{fixltx2e}
\usepackage{tcolorbox,bbold}

\newcommand{\bc}{\begin{center}}
\newcommand{\ec}{\end{center}}
\newcommand{\bt}{\begin{tabular}}
\newcommand{\et}{\end{tabular}}
\newcommand{\bi}{\begin{itemize}}
\newcommand{\ei}{\end{itemize}}
\newcommand{\be}{\begin{enumerate}}
\newcommand{\ee}{\end{enumerate}}
\newcommand{\beq}{\begin{equation}}
\newcommand{\eeq}{\end{equation}}
\newcommand{\comments}[1]{}

\begin{document}

\title{Optimal Virtualization Framework for Cellular Networks with Downlink Rate Coverage Probability Constraints}
\author{
\IEEEauthorblockN{Shubhajeet~Chatterjee$^1$, Mohammad~J.~Abdel-Rahman$^{1,2}$ and~Allen~B.~MacKenzie$^1$} 
\IEEEauthorblockA{$^1$Electrical and Computer Engineering Department, Virginia Tech, Blacksburg,VA 24061, USA \\
$^2$Electrical and Energy Engineering Department, Al Hussein Technical University (HTU), Amman 11821, Jordan \\
\{shubh92, mo7ammad, mackenab\}@vt.edu}

}

\maketitle
\begin{abstract}

Wireless network virtualization is emerging as an important technology for next-generation (5G) wireless networks. A key advantage of introducing virtualization in cellular networks is that service providers can robustly share virtualized network resources (e.g., infrastructure and spectrum) to extend coverage, increase capacity, and reduce costs. {However, the inherent features of wireless networks, i.e., the uncertainty in user equipment (UE) locations and channel conditions impose significant challenges on virtualization and sharing of the network resources.}
In this context, we propose a stochastic optimization-based virtualization framework that enables robust sharing of network resources. Our proposed scheme aims at probabilistically guaranteeing UEs' Quality of Service (QoS) demand satisfaction, while minimizing the cost for service providers, with reasonable computational complexity and affordable network overhead. 

\end{abstract}

\begin{IEEEkeywords}
Wireless network virtualization, resource allocation, rate coverage probability, chance-constrained stochastic optimization.
\end{IEEEkeywords}

\IEEEpeerreviewmaketitle

\section{Introduction}
In cellular networks, mobile network operators (MNOs) have been sharing resources (e.g., infrastructure and spectrum) as a solution to extend coverage, increase capacity, and decrease operational expenditures (OPEX) and capital expenditures (CAPEX). Recently, due to the advent of 5G with enormous coverage and capacity demands, scarcity of the overall spectrum, and potential revenue losses due to over-provisioning to serve peak demands, the motivation for sharing and virtualization has significantly increased in cellular networks. Through virtualization, wireless services can be decoupled from the network resources so that various services can efficiently share the resources. Our work provides a virtualization framework that enables network-wide robust resource sharing with reasonable computational complexity and affordable network overhead. 

\begin{figure}
\centering
\includegraphics[width=0.87\columnwidth]{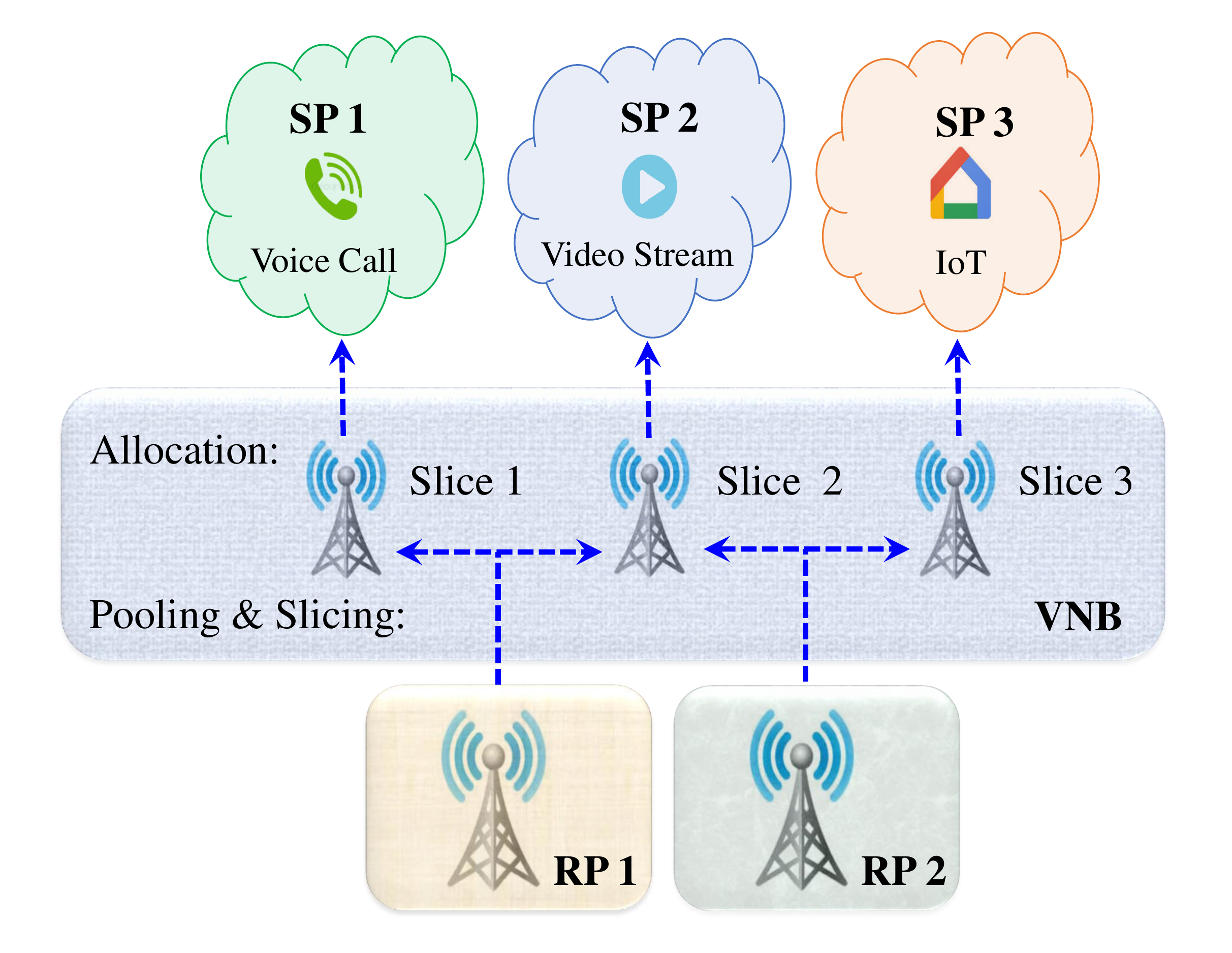}
\caption{\small{Wireless network virtualization architecture.}}
\label{figa}
\vspace{-0.25 in}
\end{figure}

We consider a three-layered architecture for wireless network virtualization (WNV) shown in Figure~\ref{figa}. The functions of each layer are described as follows~\cite{doyle2014spectrum}. Service providers (SPs) are in charge of providing regular data, voice and messaging services, as well as specialized services that apply to specific applications such as the Internet of Things (IoTs) or other current over-the-top (OTT) services. Resource providers (RPs) own the network resources. Virtual network builders (VNBs) aggregate (\textit{pool}) the resources from various RPs, create logical partitions (\textit{slice}) among these aggregated resources and allocate them to SPs. A slice of a resource is called a \textit{virtual resource} of the SP and a network built with the virtual resources is known as \textit{virtual network} of the SP. Nevertheless, to fully utilize virtualization, the virtual resource allocation process (i.e., pooling and slicing of the resources) needs to be investigated thoroughly. 


There are various challenges in the virtual resource allocation process. First, the SPs need to efficiently express their demands to the VNB. In current literature, authors proposed that an SP expresses its demand as the aggregated demands of its user equipments (UEs)~\cite{van2017defining,chen2018wireless,chang2017energy,stefanatos2017,abdel2016dimensioning}. However, due to uncertainty in UE locations and channel conditions, satisfaction of aggregated demands cannot provide any guarantee for the individual UEs' demand satisfaction. Besides, if the SPs and the VNB interact frequently to identify the instantaneous demands of individual UEs, network overhead and computational complexity would be excessive. Furthermore, the VNB needs to satisfy the SP demands as well as maximize the resource utilization (i.e., minimize over-provisioning) in the presence of the aforementioned uncertainties. 

Our goal is to address these challenges and design an efficient virtual resource allocation mechanism. Precisely, we aim to probabilistically guarantee individual UEs demand satisfaction and maximize resource utilization, with reasonable computational complexity and affordable network overhead.
Towards achieving this goal:
\begin{itemize}
\item First, we propose a new model for characterizing SP demands. The requested virtual network of an SP is fully characterized using four parameters: the minimum data rate, minimum rate coverage probability, UE intensity, and the geographical area to be covered. 
\item Second, we propose a chance-constrained virtual resource allocation framework for cellular networks. Our objective is to maximize the utilization of the resources while probabilistically guaranteeing rate demand satisfaction of UEs. This optimization framework can be used also for other service specific design criteria (e.g., delay, jitter).
\item Third, we obtain a closed-form expression for the downlink rate coverage probability of a typical virtual network. Then, using this expression, we initially solve our proposed chance-constrained virtual resource allocation problem following a heuristic greedy approach, where virtual networks are constructed gradually for the SPs until the demands of all SPs are satisfied.
\item Fourth, after simplifying the downlink rate coverage probability expression, we derive a mixed integer linear programming reformulation  of our chance-constrained virtual resource allocation problem and solve it optimally using CPLEX, which exploits the state-of-the-art branch and bound (B\&B) techniques~\cite{boyd2007branch}.
\item Fifth, considering the possibility of the optimization model being infeasible due to lack of sufficient resources in the resource pool, we propose a prioritized virtual resource allocation mechanism where virtual networks are sequentially built for SPs based on their given priorities.
\item Finally, we numerically evaluate and compare our schemes.
\end{itemize}

The rest of the paper is organized as follows. In Section~\ref{sec:Problem}, we describe the virtual resource allocation framework and the system model. In Section~\ref{sec:resource}, we present the optimal virtual resource allocation scheme. The numerical analysis is presented and discussed in Section~\ref{sec:evaluation}. Finally, the paper is concluded in Section~\ref{sec:concl}. 

\section{System Model and Framework}\label{sec:Problem}

We consider a two-dimensional geographical area $\mathcal{A}$ that is covered by a set $\mathcal{N}$ of RPs. Each RP has a set of Base Stations (BSs) deployed in $\mathcal{A}$, and the union of these sets is denoted by $\mathcal{B}$. The location of BS $b \in \mathcal{B}$ is given by $l_b$. BS $b$ operates on bandwidth $W_b$ and transmits with a constant power 1/$\mu_b$. The cost for leasing BS $b$ is $c_{b}$.  
There exists a set $\mathcal{S} =\left\{1, 2, . . . ,S\right\}$ of SPs. Each SP wants to cover the entire geographical area $\mathcal{A}$. The UEs of SP $s\in \mathcal{S}$ are assumed to be distributed in $\mathcal{A}$ according to a homogeneous Poisson Point Process (PPP) $\phi_s$ of intensity $\lambda_s$. In the following subsection, we characterize the SP demands.

\begin{figure}
\centering
\includegraphics[width=0.67\columnwidth]{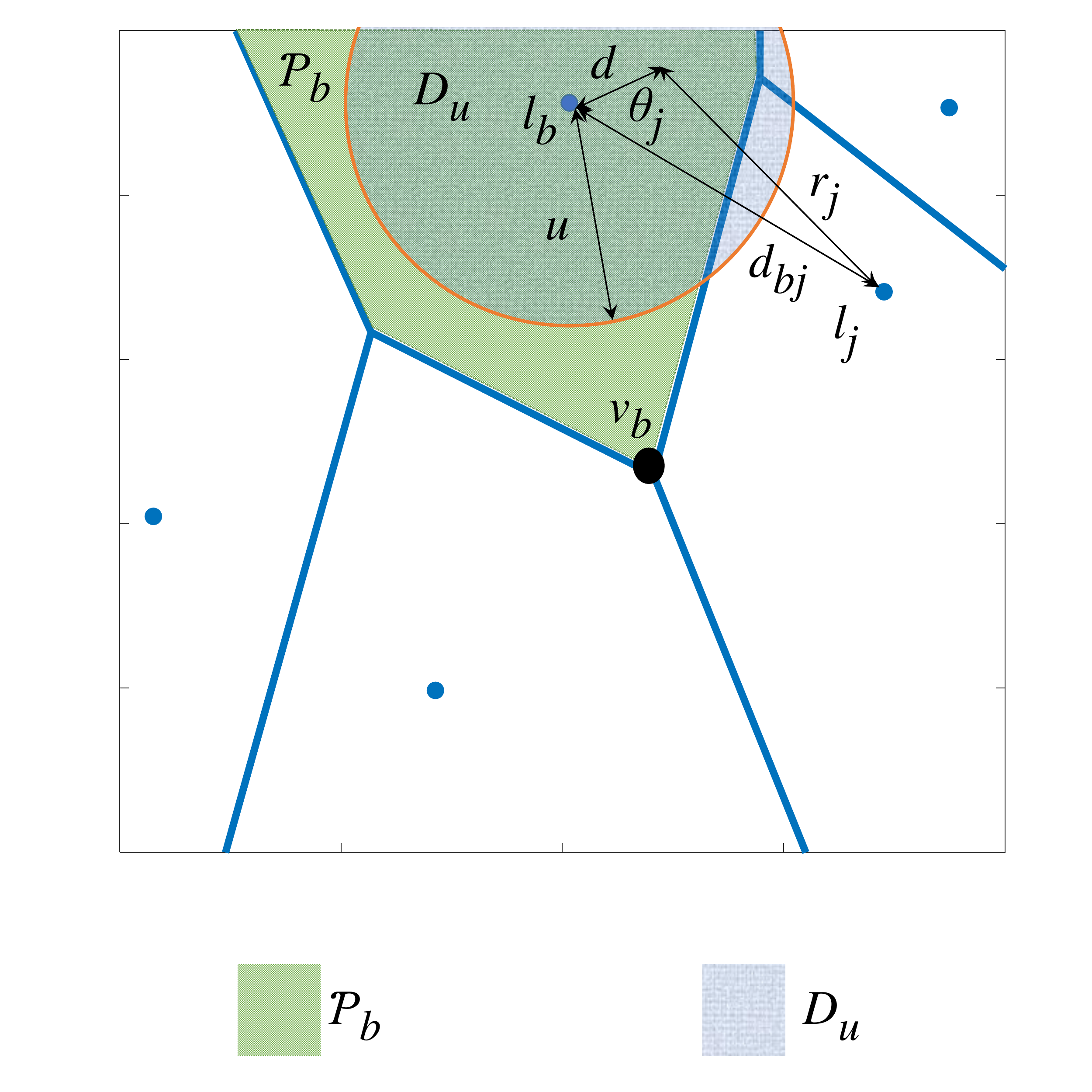}
\caption{\small{Voronoi tessellation of a set of BSs.}}
\label{Voronoi}
\end{figure}

\subsection{Demand Characterization of SPs}\label{sp demand}
SP $s, s \in \mathcal{S}$, characterizes its demand as follows: Any of its UEs located anywhere in $\mathcal{A}$ needs to have at least a data rate of $\kappa_s$ bps with a minimum probability of $\beta_s$. Let $\tilde{R}_s$ be the data rate of an arbitrarily chosen UE of SP $s$ located in $\mathcal{A}$. Then, the data rate demand of SP $s$ can be expressed as:  
\begin{equation}
\Pr \left\{ \tilde{R}_s \geq \kappa_s \right\} \geq \beta_s \nonumber.
\end{equation}

Consequently, there is a probabilistic guarantee on the demand satisfaction of the individual UEs. Let us call $\Pr \left\{ \tilde{R}_s \geq \kappa_s \right\}$ the virtual network downlink rate coverage probability. 

\subsection{Virtual Resource Allocation Framework of VNB}\label{nw model}
Upon receiving the demands from SPs, VNB leases a subset of BSs from $\mathcal{B}$ and slices them among the SPs such that their demands are satisfied. A slice of a BS provides a fraction of capacity of the BS. 

A BS can be sliced in various dimensions (e.g., time, frequency and capacity region)~\cite{van2017defining}. In this paper, we consider the VNB slices a BS in time domain. Specifically, we consider a slice of a BS is a fraction of active (or, \textit{on}) time of the BS. For example, let $\delta_{bs} \in \left[0,1\right], \; b \in \mathcal{B}, \; s \in \mathcal{S}$, be a slice of BS $b$ allocated to SP $s$. In that case, if BS $b$ serves for a time duration of $T$ then $\delta_{bs}T$ represents the time duration when SP $s$ is the only SP that accesses BS $b$. Continuing this example, $\delta_{bs}$ equals one if, SP $s$ is the only SP associated with BS $b$. Likewise, if SP $s$ is not to be associated with BS $b$ at all, then $\delta_{bs}$ would be $0$.

The slicing is implemented on a BS as follows. The VNB associates SPs with the BS and dictates the BS to reserve the fractions of its active time for each associated SPs. Now, the BS while serving the UEs of the associated SPs, ensures that each SP gets its share in each resource allocation cycle. 
Hence, in terms of slicing BSs, the VNB's job is to determine the fractions of the active time of the BSs to be allocated to SPs.

\subsection{BS Rate Allocation Model}\label{BS model}
We consider the following rate allocation model for BSs in $\mathcal{B}$. 
A UE of an SP is served by its nearest BS among the set of BSs allocated to the SP. Each BS performs a proportional rate allocation for its UEs, i.e., the rate allocated to each UE is proportional to its spectral efficiency. {Hence, assuming a saturated queue of UEs of SP $s$, the rate of a typical UE of SP $s$ associated with BS $b$ is given by}:
\begin{equation}\label{eqn:rate}
\rho = \delta_{bs} \left(\frac{W_b}{N_{bs}} \, \log_2 \left(1 + \text{SINR}_b\right)\right)
\end{equation}
where $N_{bs}$ is the total number of UEs of SP $s$ associated with BS $b$ during that time instant, $\delta_{bs}$ is the slice of BS $b$ allocated to SP $s$ and $\text{SINR}_b$ is the signal-to-interference-plus-noise ratio experienced by that typical UE.  Note that $N_{bs}$ and SINR$_b$ are stochastic variables and hence $\rho$ is a stochastic variable. 

The channel gains experienced by the UEs from their associated BS are assumed to follow a Rayleigh distribution with mean $1$ (i.e., there is no shadowing). Hence, the SINR experienced by a typical UE at an arbitrarily distance $d$ from its associated BS (say BS $b$) can be expressed as~\cite{andrews2011tractable}:
\begin{equation}
\text{SINR}_{b} = \frac{h \, d^{-\alpha}}{\sigma^2 + I}
\end{equation}
where $h$ is the stochastic channel gain, which is exponentially distributed with mean $1/\mu_b$, $\sigma^2$ is the variance of the additive noise, $\alpha$ is the pathloss exponent, and $I$ is the cumulative downlink interference from all other BSs. $I$ can be expressed as:
\begin{equation}
I = \sum_{j \in \mathcal{B}/b} g_j r^{-\alpha}_j
\end{equation} 
where $r_j$ is the distance between the typical UE and the interfering BS $j$ and $g_j$ is the stochastic gain of the channel between them. We assume that the interference also experiences Rayleigh fading without shadowing. Therefore, $g_j$ is exponentially distributed with mean $1/\mu_j$.

\subsection{Problem Statement}
{Our goal is to design a scheme to be executed at the VNB to optimally perform the virtual resource allocation, i.e., determining the optimal subset of BSs to be leased from $\mathcal{B}$ and determining the optimal fractions of active time of the leased BSs to be allocated to the SPs. Our optimality criterion is to minimize the costs of network resource aggregation (i.e., maximize the utilization of the resources) while satisfying the SP demands. Hence, we define the optimal virtual resource allocation problem as: \textit{For given demands of the SPs in $\mathcal{S}$, determine the cheapest subset of BSs to be leased from $\mathcal{B}$ such that, when sliced among SPs, these BSs can meet all SPs demand.}}

\section{Optimal Virtual Resource Allocation}\label{sec:resource}
In this section, we propose schemes that are executed at the VNB to perform the optimal virtual resource allocation. First, we formulate the problem. 
\subsection{Problem Formulation}
Let $x_b, b \in\mathcal{B}$, be a binary decision variable indicating whether to lease BS $b$ or not. $x_b$ equals one if a BS will be selected and it equals zero otherwise. 
Then, the optimal virtual resource allocation problem for the VNB can be formulated as: 

\begin{tcolorbox}[title =Problem 1: Optimal virtual resource allocation]
\vspace{-0.2in}
\begin{align}
& \underset{\left\{\substack{x_b, \delta_{bs} \\ b \in \mathcal{B}, s \in \mathcal{S}}\right\}} {\mathrm{minimize}} {\sum_{b\in \mathcal{B}}{c_{b}}\;{x_b}} \label{eqn:objective1}\\
&\hspace{-0.5 in} \text{subject to: } \nonumber\\
& \text{Pr}\left\{\tilde{R}_s \geq \kappa_s \right\} \geq \beta_s , \;\; \forall s \in \mathcal{S} \label{eqn:constraint1-1}\\
& \sum_{s \in \mathcal{S}}{\delta_{bs}} = 1, \;\; \forall b \in \mathcal{B}\label{eqn:constraint1-2}\\
& \delta_{bs} \geq 0, \;\; \forall b \in \mathcal{B}, \;\; \forall s \in \mathcal{S} \label{eqn:constraint1-3}\\
& x_b \in \{0, 1\}, \forall b \in \mathcal{B}.
\end{align}
\vspace{-0.2in}
\end{tcolorbox}

The objective function \eqref{eqn:objective1} represents the cost of the leased BSs. Constraint \eqref{eqn:constraint1-1} ensures the demand satisfaction of the SPs in $\mathcal{S}$. Constraint \eqref{eqn:constraint1-2} ensures that the utilization of the leased BSs does not exceed $100 \%$. 

In order to solve Problem 1, the key challenge is to derive a closed-form expression of constraint \eqref{eqn:constraint1-1}, i.e., the virtual network rate coverage probability obtained by the SPs.

\subsection{Virtual Network Rate Coverage Probability}

\textit{Lemma 1:} In the virtualized wireless network model described in Section II, for a set of BSs $\mathcal{B}$ the downlink rate coverage probability achieved by the virtual network of SP $s$, $s \in \mathcal{S}$, is given by \eqref{ratecoverage}. 

In \eqref{ratecoverage}, $\mathcal{P}_b$ is the region of the voronoi cell of BS $b$. $A_b$ is the area of the voronoi cell $\mathcal{P}_b$. $A$ is the area of the geographical area $\mathcal{A}$. $D_u$ is a circular disc of radius $u$ centered at $l_b$. $d$ is the distance between BS $b$ and a typical UE of SP $s$. $d_{b,j}$ is the distance between BS $b$ and an interfering BS $j$. $r_j = \sqrt{d^2 + {d^2_{b,j}} -2 d \; d_{b,j} \; \cos \theta_j}$ and $\theta_j$ is the angle between the two lines: the line connecting BS $b$ with the typical UE, and the line connecting BS $b$ with its neighboring BS $j$ as shown in Figure 2. 
    
\begin{figure*}
\rule{\textwidth}{1pt}
\begin{align}\label{ratecoverage}
\text{Pr}\left\{\tilde{R}_s \geq \kappa_s \right\}= & \sum_{b \in \mathcal{B}} \delta_{bs} \frac{A_b}{A} \Bigg[1 - e^{-\lambda A_b} \, \frac{\log 2}{2\pi W_b} \, \sum_{n=0}^{\infty} \frac{\lambda^n A_b^{n-1}}{\left(n - 1\right)!} \int^{\kappa_s}_0 2^{\frac{n \rho}{W_b}} \int^{\infty}_0 \int_0^{\left|l_b - v_b\right|} \mu_b \; u^{\alpha} \left(\sigma^2 + c\right) \exp\Bigg(-\mu_b \; u^{\alpha} \; \left(2^{\frac{n\rho}{W_b}} - 1\right) \; \nonumber \\
& \hspace{0.1in} \times \left(\sigma^2 + c\right)\Bigg) \left\{\int_{-\infty}^\infty \left\{\frac{e^{- i \omega c}}{2 \pi} \prod_{j \in \mathcal{B} \setminus b}  \int_{0}^{2 \pi} \frac{\mu_{j} \; r_{j}^{\alpha}}{\mu_{j} \; r_{j}^{\alpha} - i \omega} \; \mathrm{d}v \right\} \mathrm{d}\omega\right\} \; \frac{\mathrm{d}\left[\nabla \left\{\mathcal{P}_b \cap D_{u}\right\}\right]}{\mathrm{d}u} \; \mathrm{d}u \; \mathrm{d}c \; \mathrm{d}\rho\Bigg]. 
\end{align}
\rule{\textwidth}{1pt}
\end{figure*}

\textit{Proof :} In Theorem 1 in~\cite{OurPaper}, we have derived an expression for downlink rate coverage probability of a non-virtualized wireless network that has a similar BS rate allocation model as in Section II-C except slicing i.e., BSs allocate their full capacity (or, active time) to a single network (i.e., one SP). However, in our proposed virtualized wireless network model, a BS (say, BS $b$) is potentially time-shared among multiple SPs. Here, $\delta_{bs}, s \in \mathcal{S}$, represents the fraction of active time of BS $b$ allocated to SP $s$. In other words, at a random instant of the active time of BS $b$, SP $s$ accesses BS $b$ with a probability of $\delta_{bs}$. Therefore, the downlink rate coverage probability achieved by the virtual network of SP $s$ is obtained by multiplying the access probability $\delta_{bs}, \forall b \in \mathcal{B}$, with the downlink rate coverage probability of the non-virtualized network. Hence, we obtain the result in \eqref{ratecoverage}. 

\subsection{Solution Approach}
In this subsection, we discuss how to solve Problem 1 using \eqref{ratecoverage}. 
As can be seen, Problem 1 is a combinatorial optimization problem with NP complexity. Therefore, a simplistic solution approach is to design a heuristic greedy search algorithm where the solution is determined by iteratively adding BSs and checking the virtual network rate coverage probability constraint of the SPs from \eqref{ratecoverage}. This is a simple method to ensure demand satisfaction of the SPs. However, such a heuristic approach does not provide any theoretical guarantee for the optimality, i.e., the cost minimization. Hence, we design an efficient solution approach that can guarantee the desired optimality. 

Our solution approach is based on using CPLEX to exploit the state-of-the-art branch and bound techniques implemented in it~\cite{boyd2007branch}. To be able to exploit branch and bound (B\&B) method, we need to reformulate Problem 1 as a mixed integer linear program (MILP). Specifically, we need to express \eqref{ratecoverage} as a linear function of the binary decision variables $x_b, b \in \mathcal{B}$ and the continuous decision variables $\delta_{bs}, b \in \mathcal{B}, s \in \mathcal{S}$.
To achieve this, we want to make the following approximations and modifications over the virtual network rate coverage probability derivation in \eqref{ratecoverage}.

Recall that \eqref{ratecoverage} is derived from Theorem 1 in~\cite{OurPaper}. Therefore, we start with modifying the steps we followed in~\cite{OurPaper} to derive Theorem 1. For ease of explanation, let us briefly describe these steps before discussing the modifications. First, we derived the probability density function (PDF) of the distance of a typical UE from its nearest BS, denoted by $f_d (u)$, by assuming the coverage regions of the BSs to form vornoi cells. Then, we derived the PDF of the cumulative interference , denoted by $I$, experienced by a typical UE located at distance $d$ from its associated BS $b$, denoted by $f_I\left(c|d\right)$. Finally, based on $f_I\left(c|d\right)$ and $f_d (u)$, we derived the PDF of the received SINR and the rate coverage probability.



Note that we derived $f_d (u)$ by assuming the coverage regions of the BSs to form voronoi cells. Because $x_b, b \in \mathcal{B}$, define the shapes of the voronoi cells, expressing $f_d (u)$ as a linear function of the decision variables is extremely challenging. Instead, we will approximate the coverage region of BS $b$ as a circular area of fixed radius, denoted by $q_b$. In this case, the area of the coverage region of BS $b$ can be expressed as $\pi \; q_b^2$. In this case, the cumulative distribution function (CDF) of $d$, the distance of a typical UE from its associated BS (say $b$), can be written as:

\begin{eqnarray}\label{eqn:cdfr}
\Pr\left\{d \leq u\right\}
= \left\{
\begin{matrix}
\frac{u^2}{q_b^2}, & \text{for } 0 \leq u \leq q_b \\
1, & \text{otherwise.}
\end{matrix}
\right.
\end{eqnarray}

From the CDF~\eqref{eqn:cdfr}, the PDF of $d$ can be obtained as follows:
\begin{align}\label{eqn:pdf}
f_d(u) & = \frac{\mathrm{d}\Pr\left\{d \leq u\right\}}{\mathrm{d}u} = \left\{
\begin{matrix}
\frac{2u}{q_b^2}, & \text{for } 0 \leq u \leq q_b \\ 
0, & \text{otherwise.}
\end{matrix}
\right.
\end{align}

Following the same approximation, for a given $d$, the PDF of the angle $\theta_j$ as shown in Figure~\ref{Voronoi}, can be expressed as:
\begin{align}\label{eqn:pdf}
f_{\theta_j}(v \mid d) & = \left\{
\begin{matrix}
\frac{1}{2 \pi}, & \text{for } 0 \leq v \leq 2 \pi \\ 
0, & \text{otherwise.}
\end{matrix}
\right.
\end{align}

Next, we need to express $f_I\left(c|d\right)$ as a linear function of $x_j, \forall j \in \mathcal{B}\setminus b$. In~\cite{OurPaper}, to derive $f_I\left(c|d\right)$, we first derived its characteristic function, denoted by $\phi_I\left(\omega\right)$. Therefore, let us first modify the characteristic function $\phi_I\left(\omega\right)$. Note that if $x_j = 1$ (i.e., BS $j$ is selected) then $I_j$ (the interference caused by that BS on BS $b$) needs to be considered in $\phi_I\left(\omega\right)$. On the other hand, if $x_j = 0$, $I_j$ needs to be ignored. Hence, we express $\phi_I\left(\omega\right)$ as:
\begin{equation}
\phi_I\left(\omega\right) = \prod_{j \in \mathcal{B} \setminus b} \left(1 - x_j \left(1 - \phi_{I_j}\left(\omega\right)\right)\right)
\end{equation} 
where $\phi_{I_j}\left(\omega\right)$, the characteristic function of $I_j$, is given by:
\begin{equation}
\phi_{I_{j}}\left(\omega \right) = \frac{1}{2 \pi} \int_{0}^{2 \pi} \frac{\mu_{j} \; r_{j}^{\alpha}}{\mu_{j} \; r_{j}^{\alpha} - i \omega} \; \mathrm{d}v.
\end{equation} 

Hence, $f_I\left(c \mid d\right)$ can be expressed as a linear function of $x_j, \forall j \in \mathcal{B}\setminus b$, as: 
\begin{align}\label{Interference}
f_{I}\left(c \mid d\right) & = \frac{1}{2\pi} \int_0^\infty \Bigg\{e^{-i \omega c} \prod_{j \in \mathcal{B} \setminus b} \Big(1 - \nonumber\\
& \hspace{1in}x_j\left(1-\phi_{I_j}(\omega)\right)\Big)\Bigg\} \mathrm{d}\omega 
\end{align}

With these modifications, the PDF of the received SINR by a typical UE associated with BS $b$ can be expressed as~\eqref{snr2}.

\begin{figure*}
\rule{\textwidth}{1pt}
\begin{align}\label{snr2}
f_{\text{SINR}_b}(T) & = \int \int \! f_{\text{SINR}_b, I, d}\left(T, c, u\right) \; \mathrm{d}u \; \mathrm{d}c =  \frac{1}{2\pi} \int_0^{\infty} \int_{0}^{q_b} \mu_b \, u^{\alpha} \left(\sigma^2 + c\right) \exp\left(- \mu_b \, T \, u^{\alpha} \left(\sigma^2 + c\right)\right) \nonumber \\ 
& \hspace{1.2in}\times \left\{\int_{-\infty}^\infty \left\{e^{-i \omega c} \prod_{j \in \mathcal{B} \setminus b} \left(1 - x_j \left(1 - \frac{1}{2\pi} \int_{0}^{2\pi} \frac{\mu_j \; r_j^{\alpha}}{\mu_j \; r_j^{\alpha} - i \omega} \; \mathrm{d}v \right) \right) \right\} \; \mathrm{d}\omega \right\} \frac{2u}{{q_b}^2} \; \mathrm{d}u \; \mathrm{d}c.
\end{align}
\rule{\textwidth}{1pt}
\end{figure*}

Next, to simplify the expression of \eqref{ratecoverage}, we approximate the load of a BS for SP $s, s \in \mathcal{S}$, by the average number of UEs of SP $s$ served by that BS. Since UEs of SP $s$ are distributed according to a homogeneous PPP of intensity $\lambda_s$, the number of UEs of SP $s$ served by a BS (say, $b$) will be a Poisson random variable with parameter $\lambda_s \; \pi \; q_b^2$. Hence, the load of BS $b$ for SP $s$ is approximated by $\lambda_s \; \pi \; q_b^2$. 


With the mean load approximation, we obtain the PDF of the rate achieved by a typical UE of SP $s$, which is associated with BS $b$ (without slicing) as:
\begin{align}
f_{\text{rate}_b}(\rho) & = \left(f_{\text{SINR}_b}\left(T\right) \, \left|\frac{\mathrm{d}T}{\mathrm{d}\rho}\right|\right)_{T = \left(2^{\frac{\lambda \pi {q_b}^2 \rho }{W_b}} - 1\right)}.\label{ratePDF}
\end{align}

With slicing, as described in proof of Lemma 1, $\delta_{bs}$ represents the probability of SP $s$ accessing BS $b$ during its active time. Hence, the probability that a typical UE of SP $s$ achieves a minimum rate of $\kappa_s$ bps while being associated with BS $b$ is given by:
\begin{align}
\Pr \left\{\tilde{R}_s^b \geq \kappa_s \right\} = \delta_{bs} \; \left(1- \int_0^{\kappa_s} f_{\text{rate}_b}(\rho) \;\; \mathrm{d}\rho \right). \label{init_rateA}
\end{align}

Now, recall that UEs of an SP are assumed to be associated with the nearest BS among the set of BSs allocated to the SP. Consequently, we obtain the probability of a typical UE of SP $s$ achieves a minimum rate of $\kappa_s$ from a set of BS $\mathcal{B}$, as: 
\begin{align}
\Pr \left\{\tilde{R}_s \geq \kappa_s \right\} & = \sum_{b \in \mathcal{B}}\frac{A_b}{{A}} \Pr \left\{\tilde{R}_s^b \geq \kappa_s \right\} \nonumber\\ 
&=\sum_{b \in \mathcal{B}}\delta_{bs} \frac{A_b}{{A}} \left(1- \int_0^{\kappa_s} f_{\text{rate}_b}(\rho) \;\; \mathrm{d}\rho \right). \label{init_rate}
\end{align}

Substituting \eqref{ratePDF} in \eqref{init_rate}, we obtain a simplified expression of the downlink rate coverage probability achieved by the virtual network of SP $s$ in \eqref{ratecoverage2}.
\begin{figure*}
\rule{\textwidth}{1pt}
\begin{align}\label{ratecoverage2}
\text{Pr}\left\{\tilde{R}_s \geq \kappa_s \right\} & = \sum_{b \in \mathcal{B}} \delta_{bs} \frac{\pi {q_b}^2}{A} \Bigg\{ 1 - \frac{\lambda \log 2 }{W_b} \int_0^{\kappa_s} \left(2^{\frac{\lambda \pi {q_b}^2 \rho}{W_b}}\right)  \int_0^{\infty} \int_{0}^{q_b}  \mu_b \, u^{\alpha} \left(\sigma^2 + c\right) \exp\left(- \mu_b \, \left(2^{\frac{\lambda \pi {q_b}^2 \rho}{W_b}} - 1\right) \; u^{\alpha} \left(\sigma^2 + c\right)\right) \nonumber \\ 
& \times \left\{\int_{-\infty}^\infty \left\{e^{-i \omega c} \prod_{j \in \mathcal{B} \setminus b} \left(1 - x_j \left(1-\frac{1}{2\pi} \int_{0}^{2\pi} \frac{\mu_j \; r_j^{\alpha}}{\mu_j \; r_j^{\alpha} - i \omega} \; \mathrm{d}v \right) \right) \right\} \; \mathrm{d}\omega \right\} {u} \;  \mathrm{d}u \; \mathrm{d}c \; \mathrm{d}\rho \Bigg\}.
\end{align}


\rule{\textwidth}{1pt}
\end{figure*}


%
%

As can be seen from~\eqref{ratecoverage2}, there are two sources of non-linearity in \eqref{ratecoverage2} (with respect to decision variables $x_j \text{ and } \delta_{bs}, b \in \mathcal{B}, j \in \mathcal{B} \setminus b, s \in \mathcal{S}$). One is in the form of a product of different subsets of binary decision variables, and the other is in the form of a product of different subsets of binary and continuous decision variables. 

A product of binary decision variables, say $\prod_{j=1}^{B} x_j$, can be equivalently expressed in a linear form by (i) introducing a new auxiliary non-negative decision variable, say $x$, (ii) replace $\prod_{j=1}^{B} x_j$ by $x$, and (iii) add the following constraints:
\begin{align}
& x \leq x_j, \forall j \in \left\{1, 2, \cdots, B\right\} \nonumber\\
& x \geq \sum_{j = 1}^{B} x_j - \left(B - 1\right) \nonumber\\ \;\; 
& x \geq 0.
\end{align}

Furthermore, a product of binary and continuous decision variables, say $x_j \; \delta _{bs}$, can be equivalently expressed in a linear form by (i) introducing a new auxiliary non-negative decision variable, say $z$, (ii) replace $x_j \delta _{bs}$ by $z$ and (iii) add the following constraints:
\begin{align}
& z \leq x_j, \;\; z \leq \delta_{bs}, \;\; z \geq \delta_{bs} - (1-x_j), \;\; z \geq 0. \label{reformA}
\end{align}

Finally, we need to relate the decision variables of BS selection and slicing as:
\begin{align}
& x_b = \mathbb{1}_{\left\{ \sum_{s \in \mathcal{S}} \delta_{bs} > 0\right\}} \;\; \forall b \in \mathcal{B}. \label{reformB}
\end{align}
 
In this way, we can express \eqref{ratecoverage2} as a linear function of the decision variables. Thus, we reformulate Problem 1 as an MILP.
To make it clearer, in Appendix, we derive the complete MILP reformulation of Problem 1 when $\left|\mathcal{B}\right| = 3$.
After reformulating Problem 1 as a MILP, we solve it using the B\&B techniques implemented in CPLEX.

\subsection{Special Case}
We identify Problem 1 would be infeasible when sufficient resources are not available in resource pool $\mathcal{B}$ to satisfy all SPs demand. In that case, we consider two possibilities i.e., the VNB can either partially satisfy all SPs demand or completely satisfy some SPs demand considering their priorities. To partially satisfy all SPs demand, the VNB leases all the BSs and divides their active time equally among the SPs i.e., $\delta_{bs}= \frac{1}{\left|\mathcal{S}\right|}, \;\; \forall b \in \mathcal{B}, \forall s \in \mathcal{S}$. 
On the other hand, to completely satisfy prioritized SPs demands, the VNB ranks all SPs based on their priorities and builds their virtual networks one by one sequentially according to their ranks. Let us explain this sequential virtual resource allocation process in details. Consider the VNB builds a virtual network for an SP (say SP $s,\; s \in \mathcal{S}$). Let $\alpha_b \in [0,1],\; b \in \mathcal{B}$, denotes the available fraction of capacity in BS $b$. Now, recall that there is lack of sufficient resources to satisfy all SPs demand in the first place. Therefore, while sequentially building virtual networks for the SPs, the VNB would select all the BSs from $\mathcal{B}$ eventually. In other words, the set of BSs to be leased is fixed i.e., $\mathcal{B}$. We only need to determine the slices of these BSs to be allocated to SP $s$. Taking this into account, we formulate a virtual resource allocation problem for SP $s$ as follows:
\begin{tcolorbox}[title ={Problem 2: Virtual resource allocation for SP $s, s \in \mathcal{S}$}]
\vspace{-0.2in}
\begin{align}
& \underset{\left\{\substack{\delta_{bs}, b \in \mathcal{B}}\right\}} {\mathrm{minimize}} {\sum_{b\in \mathcal{B}}{}\;{\delta_{bs}}} \label{eqn:objective2}\\
&\hspace{-0.5 in} \text{subject to: } \nonumber\\
& \text{Pr}\left\{\tilde{R}_s \geq \kappa_s \right\} \geq \beta_s \label{eqn:constraint2-1}\\
& \delta_{bs} \leq \alpha_b, \;\; \forall b \in \mathcal{B} \label{eqn:constraint2-2}\\
& \delta_{bs} \geq 0, \;\; \forall b \in \mathcal{B} \label{eqn:constraint2-3} 
\end{align}
\vspace{-0.3in}
\end{tcolorbox}

The objective function \eqref{eqn:objective2} represents the total amount of virtual resources to be allocated to SP $s$. Constraints \eqref{eqn:constraint2-1}, \eqref{eqn:constraint2-2} and \eqref{eqn:constraint2-3} have similar purposes as in Problem 1. 

The solution approach for Problem 2 is much simpler than Problem 1. Since the set of BSs is fixed, \eqref{ratecoverage} is a linear function of the decision variables $\delta_{bs}, b \in \mathcal{B}$. Thus, Problem 2 is a Linear Program (LP). We solve this LP in CPLEX.
Now, if the solution of Problem 2 is infeasible, the VNB allocates all of the remaining slices of the BSs to SP $s$. To clarify all these steps, we provide Algorithm 1. 



\begin{algorithm}
{\small
\caption{Virtual Resource Allocation}
\label{algorithm1}
\begin{algorithmic} [1] 
\STATE Input: $\mathcal{A}$, $\mathcal{B}$, $\mathcal{S}$, $\boldsymbol{\kappa}$, $\boldsymbol{\beta}$, $\boldsymbol{\lambda}$, ranks of SPs in $\mathcal{S}$
\STATE Output: $x_b^*$, ${\delta_{bs}^* , \forall b \in \mathcal{B}, \forall s \in \mathcal{S}',  \mathcal{S}' \subseteq \mathcal{S}}$
\STATE  Reformulate Problem 1 as MILP and solve 
\IF {Solution is `infeasible'}
\STATE Sort $\mathcal {S}$ according to their ranks
\STATE Initialize: $\alpha_b$$\leftarrow$$1 \;\; \forall b \in \mathcal{B}$, $a \leftarrow 1$ 
\WHILE{$\alpha_b$ $\neq 0 \;\; \forall b \in \mathcal{B}$}
\STATE $s$$\leftarrow$$\mathcal{S}\left[a\right]$ 
\STATE Set range of $\delta_{bs}$ as [$0, \alpha_b$] $\forall b \in \mathcal{B}$
\STATE Solve Problem 2 for SP $s$
\IF {Solution is `infeasible'}
\STATE $\delta_{bs}^* \leftarrow \alpha_b \;\; \forall b \in \mathcal{B}$
\STATE EXIT
\ENDIF
\STATE $\alpha_b \leftarrow 1- \delta_{bs}^* \;\; \forall b \in \mathcal{B}$, $a \leftarrow a+1$
\ENDWHILE 
\ENDIF
\end{algorithmic}
}
\end{algorithm}

In Algorithm 1, $\mathcal{S}'$ denotes the subset of the SPs whose demands are completely satisfied. As can be seen Problem 2 is executed for individual SPs sequentially as long as the resources are available in the pool. In each iteration, based on the solution of Problem 2 i.e., $\delta_{bs}^*, \; \forall b \in \mathcal{B}$, the VNB updates $\alpha_b, \; \forall b \in \mathcal{B}$.


\subsection{Discussions}
Note that in our proposed virtualization framework, the virtualizing entities i.e., the SPs, the VNB and the RPs interact among themselves at network level instead of UE level. This brings several advantages over the existing virtualization schemes where virtual resources are allocated based on instantaneous demands of individual UEs (e.g.,\cite{chen2018wireless,chang2017energy,stefanatos2017,abdel2016dimensioning}). First, the network overhead of our proposed virtualization framework is significantly lower compared to existing schemes. Furthermore, the workload of the VNB becomes reasonable. Besides, since the optimization is performed offline, the real-time performances of the virtual networks are not being affected by the complexity of the optimization schemes.

\section{Performance Evaluation}\label{sec:evaluation}
\begin{figure*}[!t]
\begin{minipage}[t]{5.7cm}
\centering
\includegraphics[width=0.7\columnwidth]{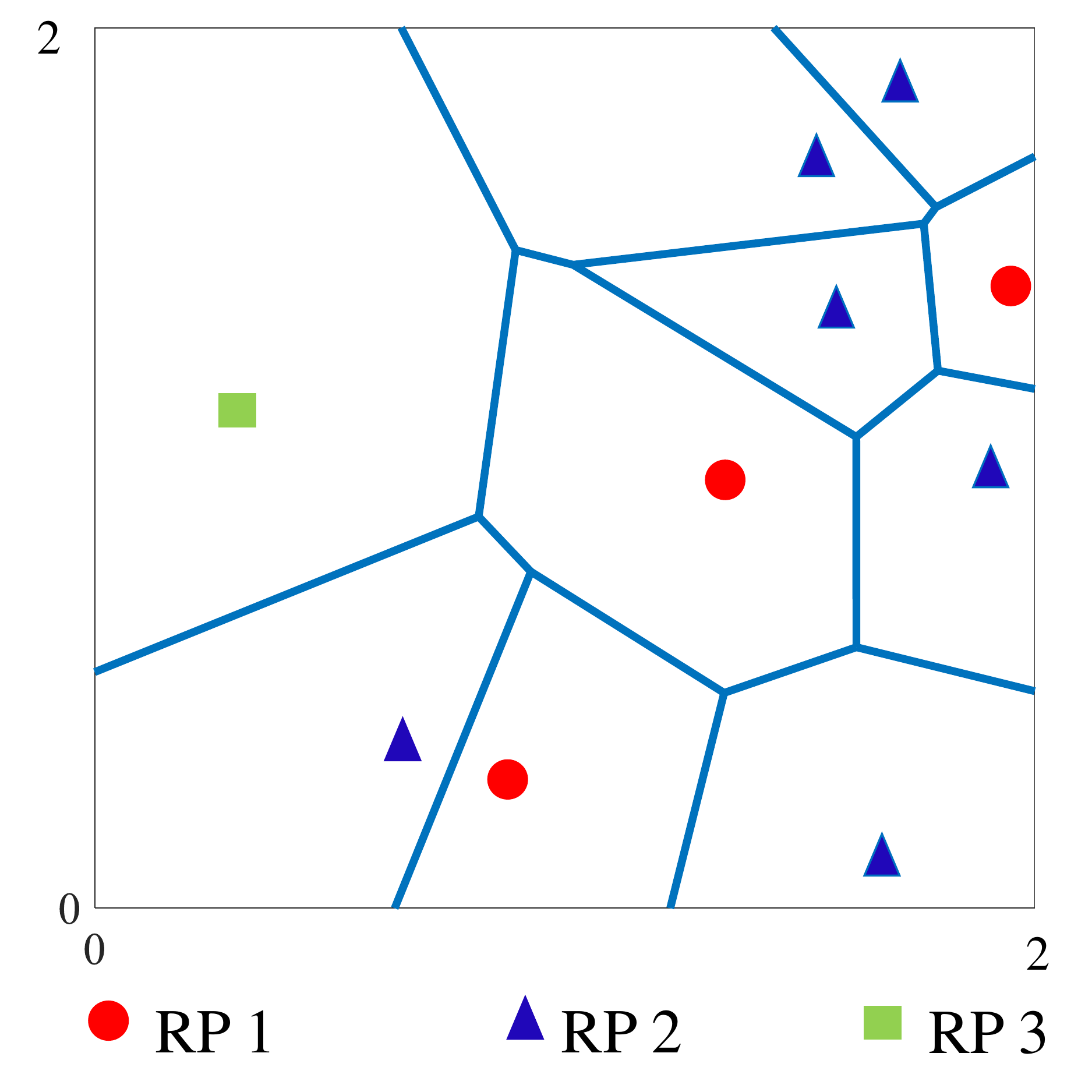}
\caption{\small{Locations of 10 BSs (Scenario I).}}
\label{voronoi}
\end{minipage}
~~
\begin{minipage}[t]{5.7cm}
\centering
\includegraphics[width=0.97\columnwidth]{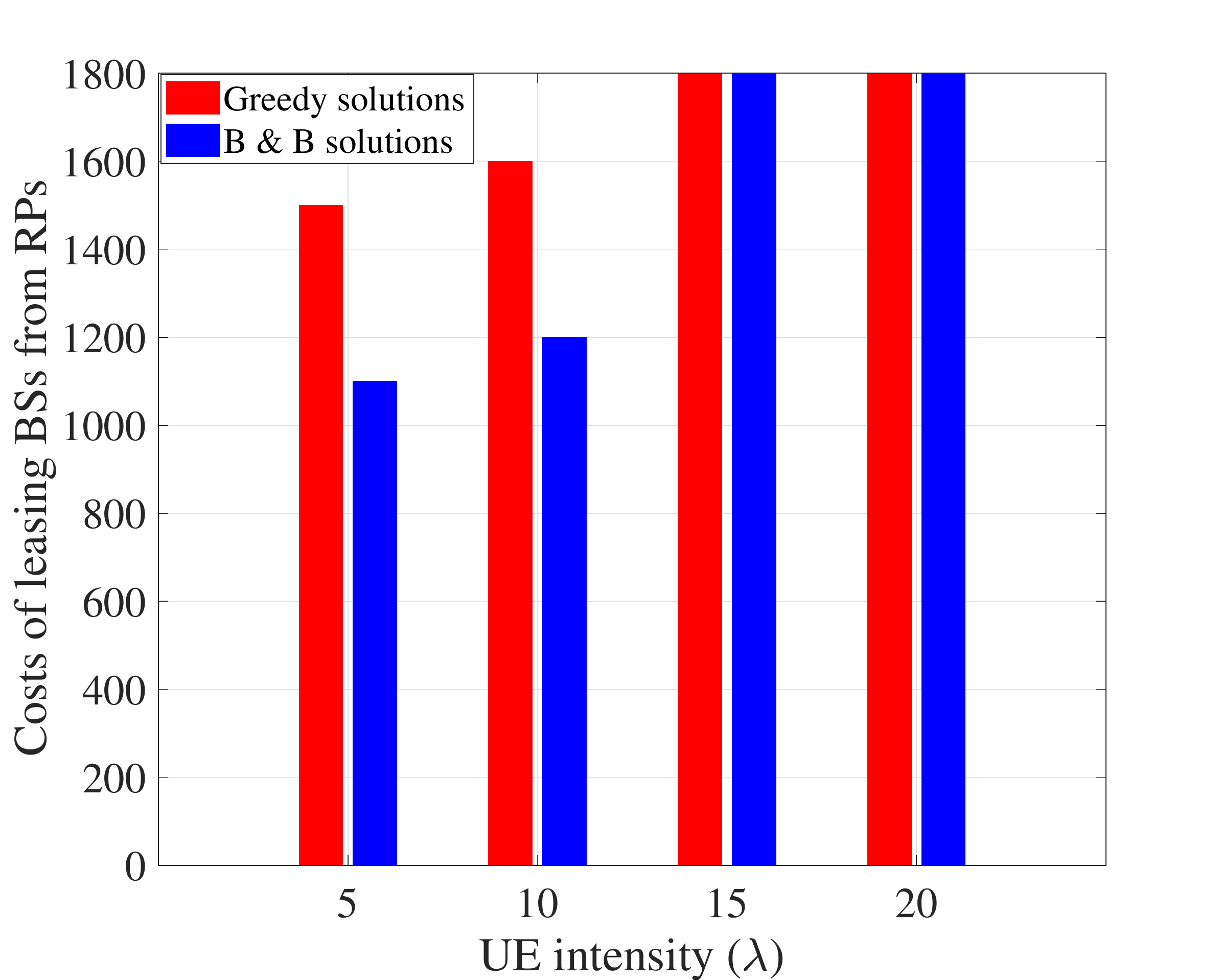}
\caption{\small{Cost of leasing BSs to satisfy the SP demands in Scenario I.}}
\label{cost}
\end{minipage}
~~
\begin{minipage}[t]{5.7cm}
\centering
\includegraphics[width=0.97\columnwidth]{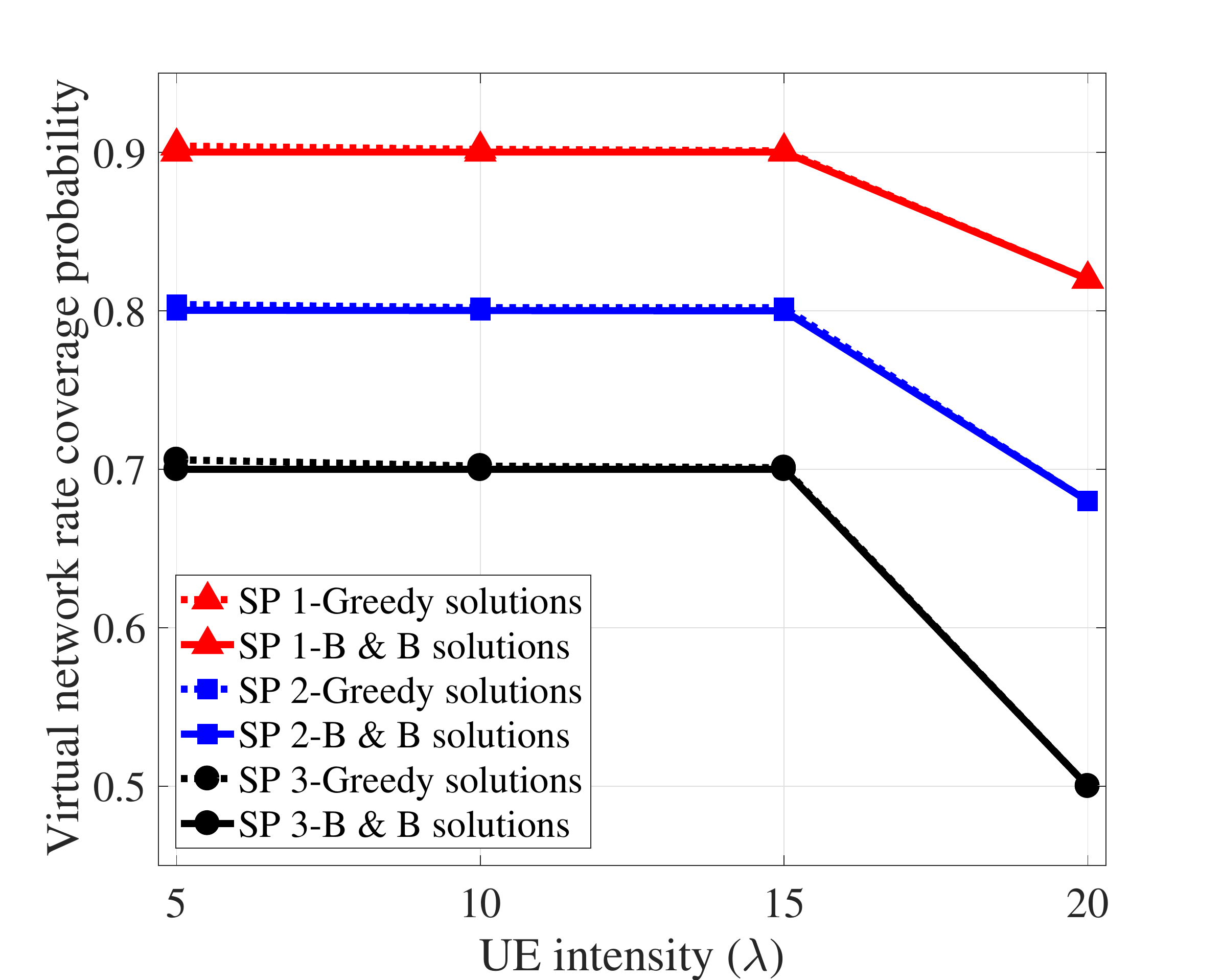}
\caption{\small{Virtual network rate coverage probability obtained by the SPs in Scenario I.}}
\label{rcp}
\end{minipage}
\end{figure*}

\begin{figure*}[!t]
\begin{minipage}[t]{5.7cm}
\centering
\includegraphics[width=0.77\columnwidth]{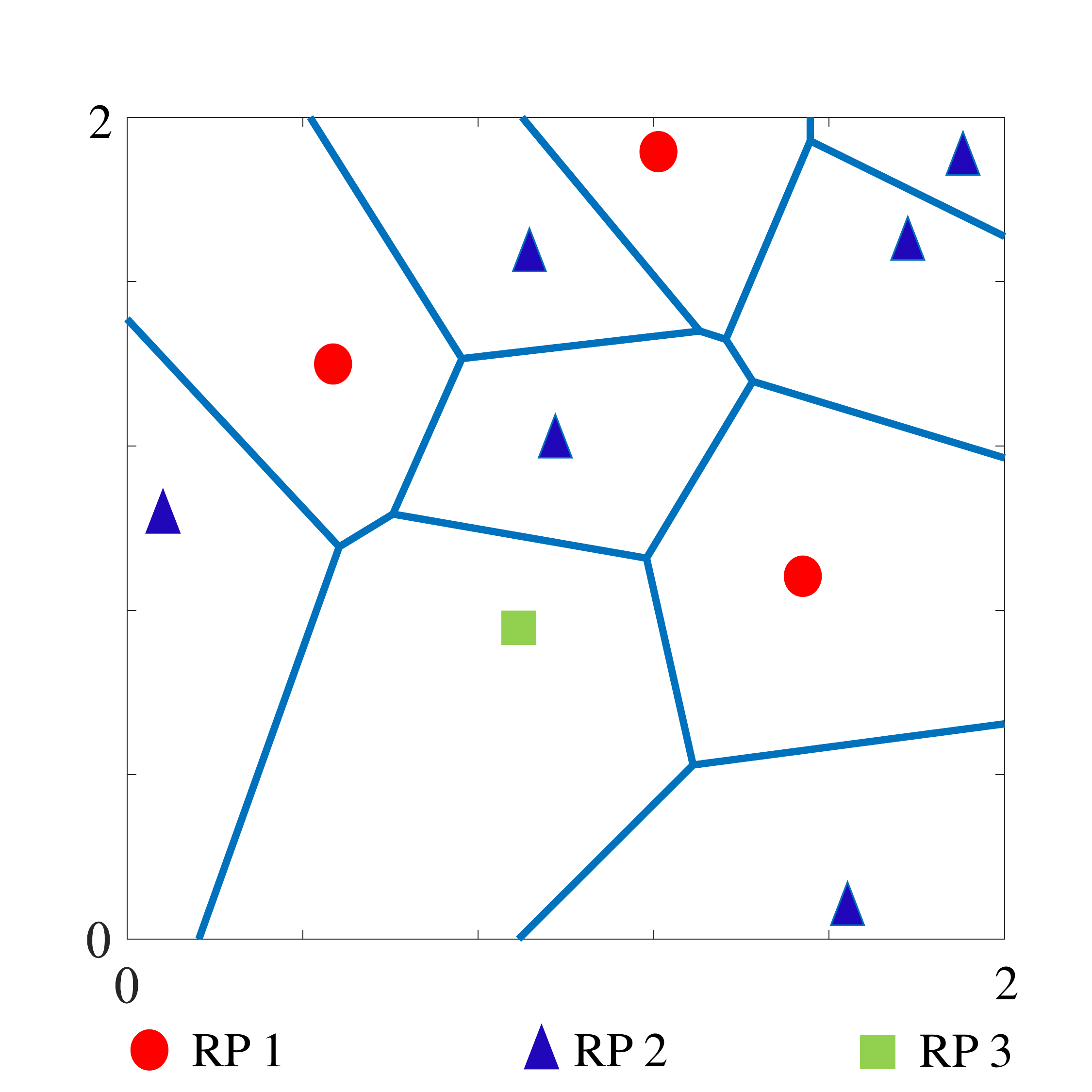}
\caption{\small{Locations of 10 BSs (Scenario II).}}
\label{bslocN}
\end{minipage}
~~
\begin{minipage}[t]{5.7cm}
\centering
\includegraphics[width=0.97\columnwidth]{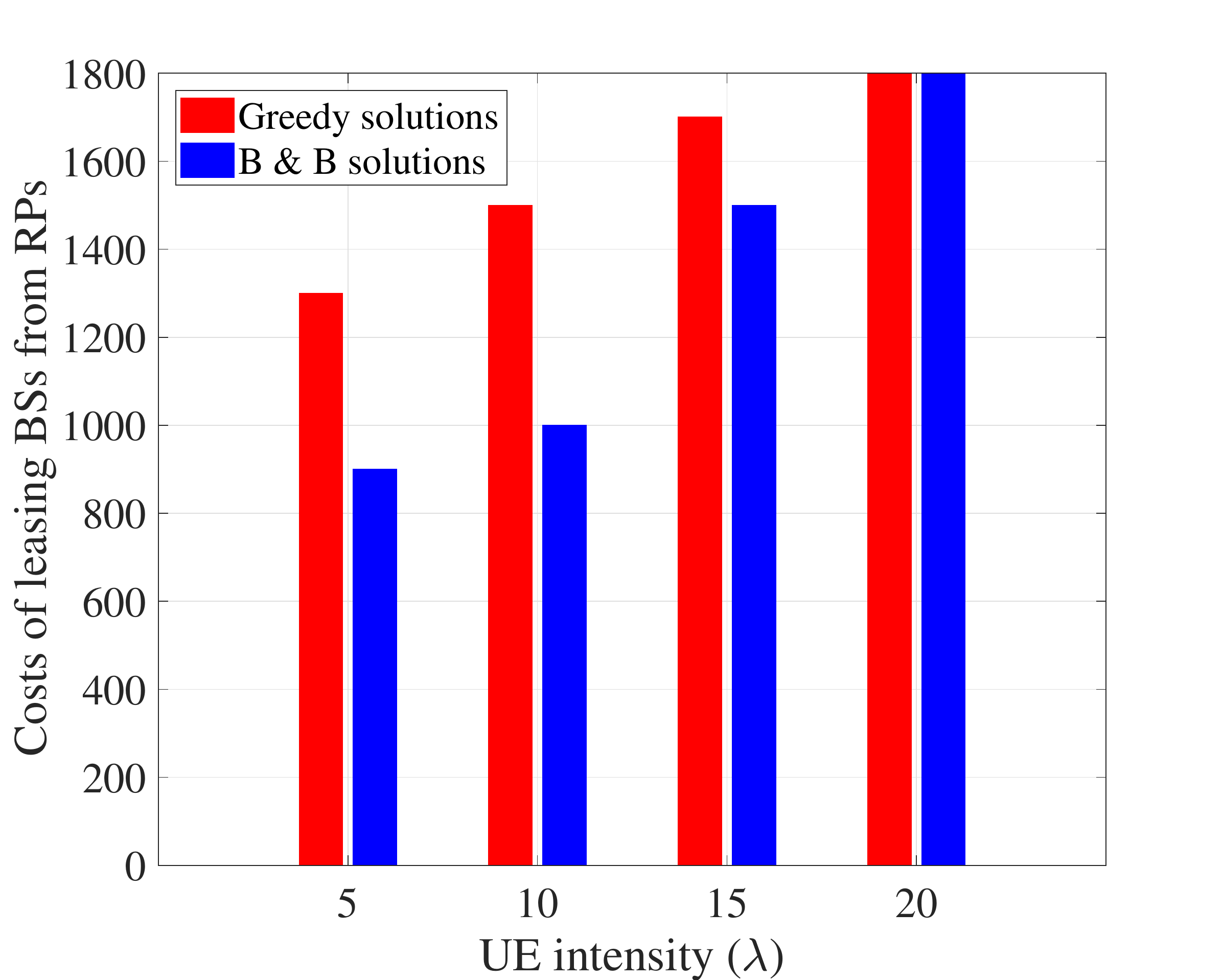}
\caption{\small{Cost of leasing BSs to satisfy the SP demands in Scenario II.}}
\label{costN}
\end{minipage}
~~
\begin{minipage}[t]{5.7cm}
\centering
\includegraphics[width=0.97\columnwidth]{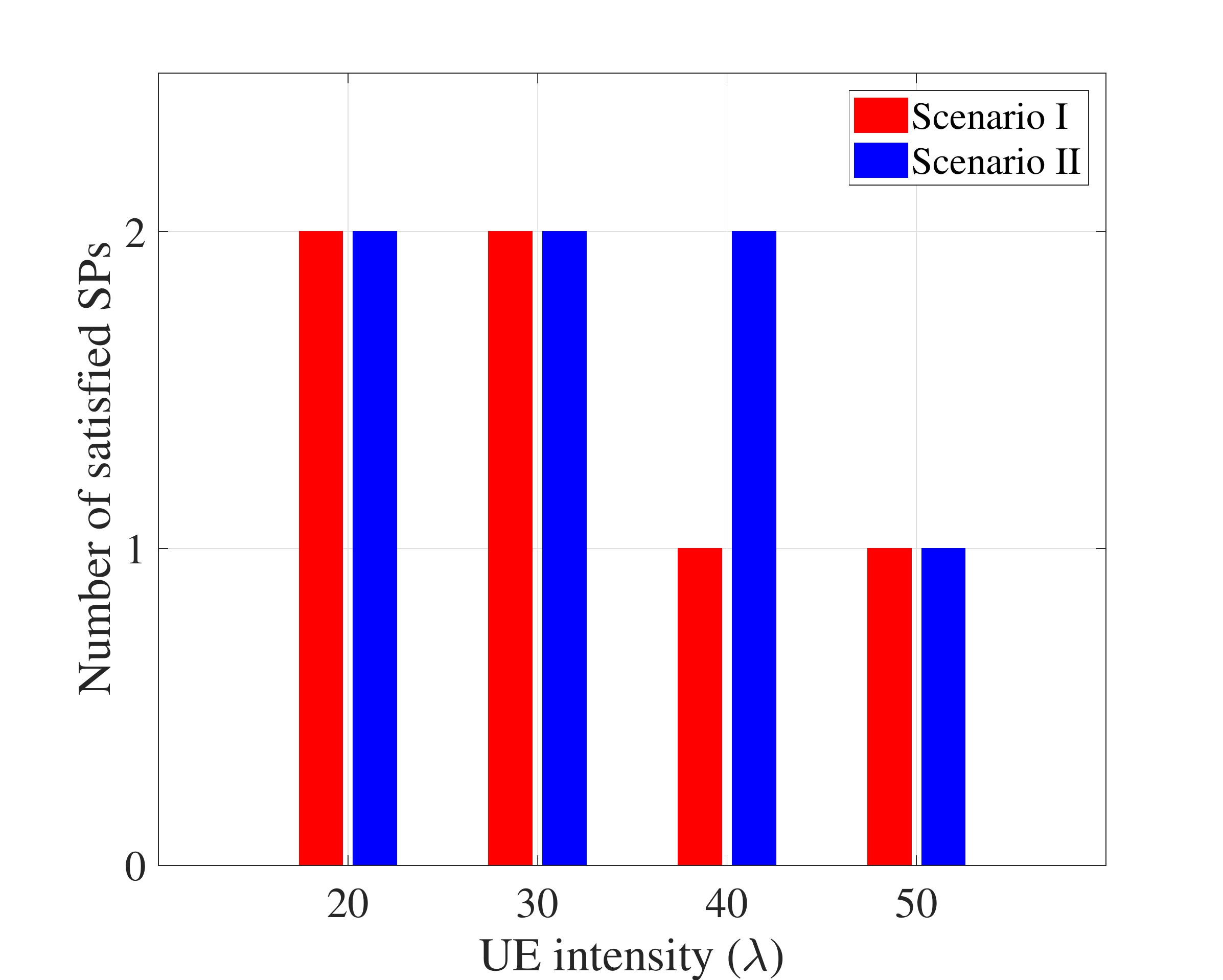}
\caption{\small{Number of satisfied SPs in ranking-based virtual resource allocation.}}
\label{prio}
\end{minipage}
\end{figure*}


In this section we evaluate the performance of our proposed virtual resource allocation scheme in terms of SPs demand satisfaction and cost minimization. 
We consider the following evaluation set up. $3$ RPs make $10$ BSs available in a geographical area of $2 \times 2$ km$^2$ as shown in Figure~\ref{voronoi}. Based on the conventional assumption that the BS locations form a homogeneous PPP, we obtain the BS locations as a realization of a homogeneous PPP of intensity $2.5/ \text{km}^2$. All $3$ BSs of RP $1$ transmit with a constant power of $23$ dBm and have cell radius of $0.3$ km (e.g., LTE small cell). All $6$ BSs of RP $2$ transmit with a constant power of $30$ dBm and have cell radius of $0.4$ km. The BS of RP $3$ transmits with a constant power of $46$ dBm and has cell radius of $1$ km (e.g., LTE macro cell). All $10$ BSs operate over a bandwidth of $20$ MHz. Noise variance ($\sigma^2$) is set to $-174$ dBm/Hz. Pathloss exponent ($\alpha$) is set to $4$. The cost of leasing a BS is $\$ 100$ from RP $1$, $ \$ 200$ from RP $2$ and $ \$ 300$ from RP $3$.
There are $3$ SPs who wish to provide wireless services within the considered geographical area. SP $1$ requires its UEs to have a minimum data rate of $512$ Kbps (i.e., required data rate for indoor IoT applications) with probabilistic guarantee of $0.9$. SP $2$ requires its UEs to have a minimum data rate of $1$ Mbps (i.e., required data rate for high definition video calling) with probabilistic guarantee of $0.8$. SP $3$ requires its UEs to have a minimum data rate of $4$ Mbps (i.e., required data rate for high definition video streaming) with probabilistic guarantee of $0.7$. Each SP has same UE intensity. 

In this set up, we vary the UE intensity of the SPs and execute Problem 1 based on our proposed solution approaches i.e., greedy-search-based heuristic approach and B\&B-techniques-based approach. For this analysis we do not prioritize any SP i.e., if the MILP is infeasible, Then $\delta_{bs} = \frac{1}{3}, \; \forall b \in \mathcal{B}, \;\forall s \in \mathcal{S}$. We plot the the costs of leasing BSs in Figure~\ref{cost} and the virtual network rate coverage probability obtained by the SPs in Figure~\ref{rcp}.  
It can be seen from Figure~\ref{rcp} when sufficient resources are made available to the VNB (in other words, as long as the UE intensity is moderate i.e., less than or equal to $6$ times the BS intensity), both of the schemes can ensure SPs demand satisfaction. However, as can be seen in Figure~\ref{cost}, the B\&B solutions are more cost efficient compare to the greedy solutions. 

Next, we analyze the impact of the locations of the BSs on the virtualized cellular network. Keeping other parameters intact, we change the locations of the BSs (i.e., consider another realization of the aforementioned homogeneous PPP of the BS distribution) as shown in Figure~\ref{bslocN}. Then, we repeat our previous evaluation of Problem 1 and plot the costs of leasing BSs in Figure~\ref{costN}. Comparing Figure~\ref{cost} and Figure~\ref{costN}, it can be seen that the locations of the BSs significantly impact on cost minimization. Finally, we evaluate the performance of the sequential  virtual resource allocation scheme in the two considered BS topologies. We assume an arbitrary ranking scheme as: SP1$>$SP2$>$SP3 and plot the number of satisfied SPs for each scenario in Figure~\ref{prio}. As can be seen, the locations of the BSs influence the number of SPs that can be satisfied. Hence, we conclude that the conventional BS-distribution-based cellular network analysis might not be viable when virtualization is introduced. To optimally perform pooling and slicing, we need to consider individual BSs.

\section{Conclusions}\label{sec:concl}
In this paper, we provided an optimal virtual resource allocation mechanism that guarantees probabilistic satisfaction for the UEs demands while minimizing the costs of leasing BSs (i.e., maximizing the utilization of the resources). We evaluated the proposed scheme and studied the gains brought by the optimal scheme compare to the heuristic scheme. 


\bibliographystyle{IEEEtran}
\bibliography{ref_shubhaV}

\twocolumn[{
 \begin{@twocolumnfalse}
   
\section {Appendix}

%
Here, we provide an illustration of mixed integer linear program (MILP) reformulation of Problem 1 when $\mathcal{B}= \left\{1,2,3\right\}$. 

For ease of representation, let
\begin{align}
& \hspace{-2.7in}g_b\left(u, c, \rho\right) = \mu_b \; u^{\alpha} \; \left(\sigma^2 + c\right) \exp\left(- \mu_b \left(2^{\frac{\lambda \pi {q_b}^2\rho}{W_b}} - 1\right) \; u^{\alpha} \; \left(\sigma^2 + c\right)\right) \; u \nonumber
\end{align}
and
\begin{align}
& \hspace{-3.5in}h_j\left(u, \omega \right) = \frac{1}{2 \pi} \int_{0}^{2\pi} \frac{\mu_j \; r_j^{\alpha}}{\mu_j \; r_j^{\alpha} - i \omega} \; \mathrm{d}v. \nonumber
\end{align} 
Then,~\eqref{ratecoverage2} can be written as~\eqref{ratecoverage3}.
%
\begin{align}\label{ratecoverage3}
&\hspace{-0.8in}\text{Pr}\left\{\tilde{R}_s \geq \kappa_s \right\} = \sum_{b \in \mathcal{B}} \delta_{bs} \frac{ \pi {q_b}^2}{A}\Bigg\{1 - \frac{\lambda \log 2}{W_b} \int_0^{\kappa_s} 2^{\frac{\lambda \pi {q_b}^2\rho}{W_b}} \int_0^{\infty} \int_{0}^{q_b} g_b\left(u, c, \rho\right) \nonumber \\
& \hspace{2in} \times \int_{-\infty}^\infty e^{-i \omega c} \prod_{j \in \mathcal{B} \setminus b} \left(1 - x_j \left(1 - h_j\left(u, \omega\right)\right)\right) \; \mathrm{d}\omega \; \mathrm{d}u \; \mathrm{d}c \; \mathrm{d}\rho \Bigg\}. 
\end{align}

Now, we expand \eqref{ratecoverage3} with respect to $x_b, \;\; \delta_{bs}, \;\; b \in \mathcal{B}, s \in \mathcal{S}$, as follows:
\begin{align}  
\text{Pr}\left\{\tilde{R}_s \geq \kappa_s \right\}  & = \left[\sum_{b\in \mathcal{B}}\delta_{bs}\frac{ \pi {q_b}^2}{A}\right] - \frac{\lambda \log 2}{A}\Bigg[\frac{\delta_{1s} \pi {q_1}^2}{W_1} \int_0^{\kappa_s} 2^{\frac{\lambda \pi {q_1}^2\rho}{W_1}} \int_0^{\infty} \int_{0}^{q_1} g_1\left(u, c, \rho\right) \int_{-\infty}^\infty e^{-i \omega c} \left(1 - x_2\left(1 - h_2\left(u, \omega\right)\right)\right) \nonumber \\
& \hspace{3in} \times \left(1 - x_3\left(1 - h_3\left(u, \omega\right)\right)\right) \; \mathrm{d}\omega \; \mathrm{d}u \; \mathrm{d}c \; \mathrm{d}\rho \nonumber \\
& \hspace{1.5in} + \frac{\delta_{2s} \pi {q_2}^2}{W_2} \int_0^{\kappa_s} 2^{\frac{\lambda \pi {q_2}^2\rho}{W_2}} \int_0^{\infty} \int_{0}^{q_2} g_2\left(u, c, \rho\right) \int_{-\infty}^\infty e^{-i \omega c} \left(1 - x_1\left(1 - h_1\left(u, \omega\right)\right)\right) \nonumber \\
& \hspace{3in} \times \left(1 - x_3\left(1 - h_3\left(u, \omega\right)\right)\right) \; \mathrm{d}\omega \; \mathrm{d}u \; \mathrm{d}c \; \mathrm{d}\rho \nonumber \\
& \hspace{1.5in} + \frac{\delta_{3s} \pi {q_3}^2}{W_3} \int_0^{\kappa_s} 2^{\frac{\lambda \pi {q_3}^2\rho}{W_3}} \int_0^{\infty} \int_{0}^{q_3} g_3\left(u, c, \rho\right) \int_{-\infty}^\infty e^{-i \omega c} \left(1 - x_1 \left(1 - h_1\left(u, \omega \right) \right) \right) \nonumber \\
& \hspace{3in} \times \left(1 - x_2\left(1 - h_2\left(u, \omega\right)\right)\right) \; \mathrm{d}\omega \; \mathrm{d}u \; \mathrm{d}c \; \mathrm{d}\rho \Bigg] \nonumber \\
& = \left[\sum_{b\in \mathcal{B}}\delta_{bs}\frac{ \pi {q_b}^2}{A}\right] - \frac{\lambda \log 2}{A}\Bigg[\frac{\delta_{1s} \pi {q_1}^2}{W_1} \int_0^{\kappa_s} 2^{\frac{\pi {q_1}^2 \rho}{W_1}} \int_0^{\infty} \int_{0}^{q_1} g_1\left(u, c, \rho\right) \int_{-\infty}^\infty e^{-i \omega c} \Bigg(1 - \sum_{\substack{b \in \mathcal{B} \\ b \neq 1}} x_b + \sum_{\substack{b \in \mathcal{B} \\ b \neq 1}} x_b \; h_b\left(u, \omega\right) \nonumber \\
& \hspace{2in} + \prod_{\substack{b \in \mathcal{B} \\ b \neq 1}} x_b \Bigg(1 - \sum_{\substack{b \in \mathcal{B} \\ b \neq 1}} h_b\left(u, \omega\right) + \prod_{\substack{b \in \mathcal{B} \\ b \neq 1}} h_b\left(u, \omega\right)\Bigg)\Bigg) \; \mathrm{d}\omega \; \mathrm{d}u \; \mathrm{d}c \; \mathrm{d}\rho \nonumber \\
& \hspace{1.2in} + \frac{\delta_{2s} \pi {q_2}^2}{W_2} \int_0^{\kappa_s} 2^{\frac{\pi {q_2}^2 \rho}{W_2}} \int_0^{\infty} \int_{0}^{q_2} g_2\left(u, c, \rho\right) \int_{-\infty}^\infty e^{-i \omega c} \Bigg(1 - \sum_{\substack{b \in \mathcal{B} \\ b \neq 2}} x_b + \sum_{\substack{b \in \mathcal{B} \\ b \neq 2}} x_b \; h_b\left(u, \omega\right) \nonumber \\
& \hspace{2in} + \prod_{\substack{b \in \mathcal{B} \\ b \neq 2}} x_b \Bigg(1 - \sum_{\substack{b \in \mathcal{B} \\ b \neq 2}} h_b\left(u, \omega\right) + \prod_{\substack{b \in \mathcal{B} \\ b \neq 2}} h_b\left(u, \omega\right)\Bigg)\Bigg) \; \mathrm{d}\omega \; \mathrm{d}u \; \mathrm{d}c \; \mathrm{d}\rho \nonumber \\
& \hspace{1.2in} + \frac{\delta_{3s} \pi {q_3}^2 }{W_3} \int_0^{\kappa_s} 2^{\frac{\pi {q_3}^2 \rho}{W_3}} \int_0^{\infty} \int_{0}^{q_3} g_3\left(u, c, \rho\right) \int_{-\infty}^\infty e^{-i \omega c} \Bigg(1 - \sum_{\substack{b \in \mathcal{B} \\ b \neq 3}} x_b + \sum_{\substack{b \in \mathcal{B} \\ b \neq 3}} x_b \; h_b\left(u, \omega\right) \nonumber \\
& \hspace{2in} + \prod_{\substack{b \in \mathcal{B} \\ b \neq 3}} x_b \Bigg(1 - \sum_{\substack{b \in \mathcal{B} \\ b \neq 3}} h_b\left(u, \omega\right) + \prod_{\substack{b \in \mathcal{B} \\ b \neq 3}} h_b\left(u, \omega\right)\Bigg)\Bigg) \; \mathrm{d}\omega \; \mathrm{d}u \; \mathrm{d}c \; \mathrm{d}\rho\Bigg]. \label{rate3}
\end{align} 

\end{@twocolumnfalse}
}]

\begin{figure*}

Hence, from \eqref{reformA} and \eqref{reformB}, we reformulate Problem 1 as follows:
\begin{tcolorbox}[title=Problem 3: MILP reformulation of the optimal virtual resource allocation problem]
\begin{align} 
& \underset{\left\{\substack{x_{b} \in \mathcal{B}}\right\}}{\mathrm{minimize}} \; \sum_{b \in \mathcal{B}} \; c_b \;\; x_{b} \\
\text{subject to:} \nonumber \\
& \left\{\sum_{b=1}^3 \delta_{bs} -\sum_{b=1}^3 U_b \delta_{bs} + \sum_{\substack{b,j \in \mathcal{B} \\ j \neq b}} U_b z_{bjs} + \sum_{\substack{b,j \in \mathcal{B} \\ j \neq b}} V_{bj} z_{bjs} - z_{231s}Y_1 - z_{132s} Y_2 - z_{123s} Y_3\right\} \geq \beta_s, \;\; \forall s \in \mathcal{S} \\
& z_{12s} \leq x_2, \;\; z_{12s} \leq \delta_{1s}, \;\; z_{12s} \geq \delta_{1s}- (1-x_2), \;\; z_{12s} \geq 0, \;\; \;\; \forall s \in \mathcal{S}\\  
& z_{13s} \leq x_3, \;\; z_{13s} \leq \delta_{1s}, \;\; z_{13s} \geq \delta_{1s}- (1-x_3), \;\; z_{13s} \geq 0, \;\; \;\; \forall s \in \mathcal{S}\\
& z_{21s} \leq x_1, \;\; z_{21s} \leq \delta_{2s}, \;\; z_{21s} \geq \delta_{2s}- (1-x_1), \;\; z_{21s} \geq 0, \;\; \;\; \forall s \in \mathcal{S}\\ 
& z_{23s} \leq x_3, \;\; z_{23s} \leq \delta_{2s}, \;\; z_{23s} \geq \delta_{2s}- (1-x_3), \;\; z_{21s} \geq 0, \;\; \;\; \forall s \in \mathcal{S}\\ 
& z_{31s} \leq x_1, \;\; z_{31s} \leq \delta_{3s}, \;\; z_{31s} \geq \delta_{3s}- (1-x_1), \;\; z_{31s} \geq 0, \;\; \;\; \forall s \in \mathcal{S}\\   
& z_{32s} \leq x_2, \;\; z_{31s} \leq \delta_{3s}, \;\; z_{32s} \geq \delta_{3s}- (1-x_2), \;\; z_{32s} \geq 0, \;\; \;\; \forall s \in \mathcal{S}\\   
& x_{12} \leq x_1, \;\; x_{12} \leq x_2, \;\; x_{12} \geq x_1 + x_2 - 1, \;\; x_{12} \geq 0 \\ 
& x_{13} \leq x_1, \;\; x_{13} \leq x_3, \;\; x_{13} \geq x_1 + x_3 - 1, \;\; x_{13} \geq 0 \\ 
& x_{23} \leq x_2, \;\;  x_{23} \leq x_3, \;\; x_{23} \geq x_2 + x_3 - 1, \;\; x_{23} \geq 0 \\
& z_{231s} \leq x_{23}, \;\; z_{231s} \leq \delta_{1s}, \;\; z_{231s} \geq \delta_{1s}- (1-x_{23}), \;\; z_{231s} \geq 0, \;\; \;\; \forall s \in \mathcal{S}\\ 
& z_{123s} \leq x_{12}, \;\; z_{123s} \leq \delta_{3s}, \;\; z_{123s} \geq \delta_{3s}- (1-x_{123}), \;\; z_{123s} \geq 0, \;\; \;\; \forall s \in \mathcal{S}\\ 
& z_{132s} \leq x_{13}, \;\; z_{132s} \leq \delta_{2s}, \;\; z_{132s} \geq \delta_{2s}- (1-x_{132}), \;\; z_{132s} \geq 0, \;\; \;\; \forall s \in \mathcal{S}\\
& \sum_{s \in \mathcal{S}}{\delta_{bs}} = 1, \;\; \forall b \in \mathcal{B}\\
& x_b = \mathbb{1}_{\left\{ \sum_{s \in \mathcal{S}} \delta_{bs} > 0\right\}} ,\;\; \forall b \in \mathcal{B} \\
& \delta_{bs} \geq 0, \;\; \forall b \in \mathcal{B}, \;\; \forall s \in \mathcal{S}
\end{align}
\end{tcolorbox}
\end{figure*}


\begin{figure*}
where,
\begin {align}
& U_b = \frac{\lambda \pi {q_b}^2  \log 2}{W_b A} \int_0^{\kappa_s} 2^{\frac{\pi {q_b}^2 \rho}{W_b}} \int_0^{\infty} \int_{0}^{q_b} g_b \left(u,c,\rho \right) \left\{\int_{-\infty}^\infty e^{-i \omega c} \; \mathrm{d}\omega \right\} \,  \mathrm{d}u \, \mathrm{d}c \, \mathrm{d}\rho \;\;  \nonumber\\
& V_{bj}= \frac{\lambda \pi {q_b}^2  \log 2}{W_b A} \int_0^{\kappa_s} 2^{\frac{\pi {q_b}^2 \rho}{W_b}} \int_0^{\infty} \int_{0}^{q_b} g_b \left(u,c,\rho \right) \left\{\int_{-\infty}^\infty e^{-i \omega c} h_j\left(u, \omega \right) \; \mathrm{d}\omega\right\} \;  \mathrm{d}u \; \mathrm{d}c \; \mathrm{d}\rho \nonumber\\
& Y_b= \frac{\lambda \pi {q_b}^2  \log 2}{W_b A} \int_0^{\kappa_s} 2^{\frac{\pi {q_b}^2 \rho}{W_b}} \int_0^{\infty} \int_{0}^{q_b} g_b \left(u,c,\rho \right) \Bigg\{ \int_{-\infty}^\infty e^{-i \omega c} (1- \nonumber\\
&\hspace{2.7in} \sum_{\substack{j \in \mathcal{B} \\ j \neq b}} h_j\left(u, \omega\right) + \prod_{\substack{j \in \mathcal{B} \\ j \neq b}} h_j\left(u, \omega\right)) \; \mathrm{d}\omega \Bigg\} \,  \mathrm{d}u \; \mathrm{d}c \; \mathrm{d}\rho \;\; \;\; \forall b,j \in \mathcal{B}, \;\; j \neq b. \nonumber
\end{align}
Note that in Problem 3, the functions $U_b, V_{bj}$, $Y_1$, $Y_2$ and $Y_3$ do not contain any of the decision variables $x_b, \delta_{bs}, b \in \mathcal{B}, s \in \mathcal{S}$. Consequently, Problem 3 is a MILP reformulation of Problem 1.
\end{figure*}



\end{document}